# Optical Absorption and Photocurrent Spectroscopy for Investigating Important Length Scales in Molecular Electronics

V Mukundan[1], J A Fereiro[1], R L McCreery[1], and A J Bergren[*2]

[1]Department of Chemistry, University of Alberta, Edmonton, AB, Canada

[2]National Institute for Nanotechnology, National Research Council Canada, 11421 Saskatchewan Dr, Edmonton, AB Canada T6G 2M9

*Correspondence to: adam.bergren@nrc.ca

Abstract: We report the optical absorption characteristics and the photocurrent spectra for large-area molecular junctions containing different molecular structures and thicknesses. Through experimental assessment of the optical absorption of molecules in solution and chemisorbed onto transparent, conductive carbon films, the energy ranges and intensity of absorption for a variety of molecular layers as a function of thickness were determined. Our findings show that optical absorption is red-shifted for a chemisorbed molecular layer relative to the molecule in solution, indicating electronic coupling between the molecule and the surface. However, we also find that from 2-22 nm in thickness, the absorption characteristics follow Beer's law, with no further shifts in the wavelength of maximum absorption or low-energy absorption onset. For energies where no absorption occurs, an internal photoemission process takes place resulting in photocurrent that can be sustained over a limited range. For energies at which the molecular component absorbs light, photocurrent can proceed over longer distances, and the trends in photocurrent yield versus thickness shows interesting trends that may be related to the distance scale over which charge carriers can communicate with the conductive contacts. This method is particularly relevant for strongly coupled systems that rely on molecular properties to mediate charge transport, since excitation takes place directly within the molecular layer, and provides a direct probe of energetics in systems that show bulk-limited transport.

## 1 INTRODUCTION

Molecular electronics is a field of study that seeks to understand electronic transport across very thin layers of organic molecules, typically under ~10 nm, with a few studies extending into thicker films[1, 2]. The molecular junction (see Figure 1A) can contain a single molecule or many molecules (example structures in Figure 1B) in parallel between two conductive (or semi-conductive) contacts, and is the basic unit studied in molecular electronics[3]. In any molecular electronic system, the distance over which charge carriers are transported is of interest since it can determine the charge transport mechanisms that dominate the electronic properties of the device. A fundamental understanding of the relationship between molecular structure (including thickness) and the electronic function of nanoscale electronic devices that incorporate them has been a key goal in molecular electronics[4-8]. To this end, methods for investigating important lengths scales can provide critical insights into device behaviour.

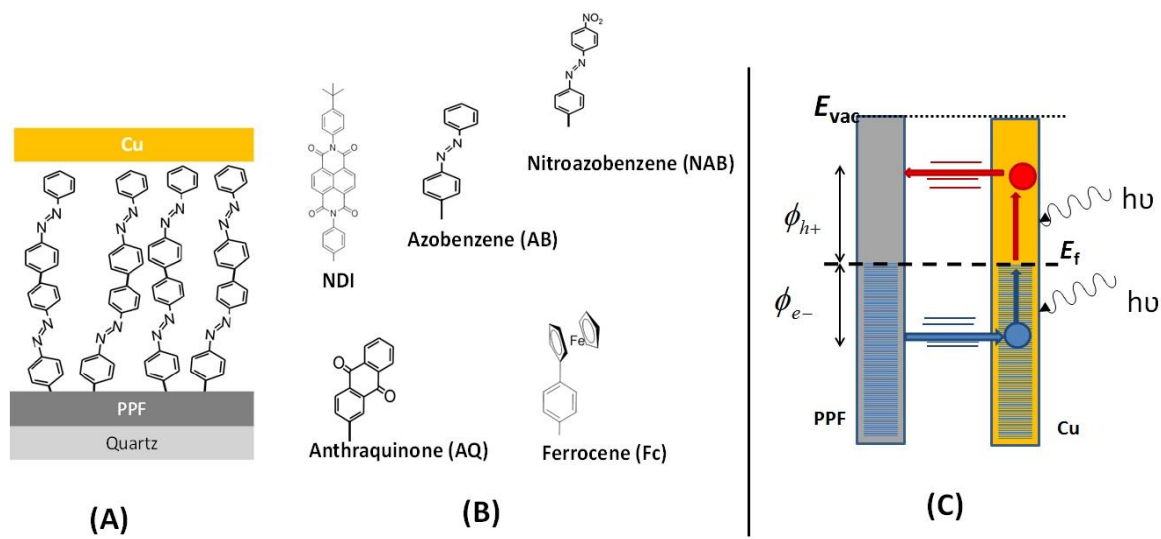

**Figure 1**. (A) Illustration of the molecular junction structure used in this study, with an oriented but disordered molecular layer covalently bonded to a carbon contact (PPF, or pyrolyzed photoresist film) with a 20 nm layer of Cu as the top contact. (B) Structures and abbreviations of molecules used in to construct devices. (C) Energy level diagram of internal photoemission.

Techniques to evaluate charge transport, molecular structure, interfacial energetics, and other aspects of molecular electronic devices have expanded as the quest to design specific electronic function has continued[9]. From a variety of studies in molecular electronics, a few of the "rules" that dictate charge transport have been determined, although the exact situation depends on many of the details of the specific device under study. For example, the basic charge transport mechanism in operation can vary, depending at least on the thickness, applied voltage (and/or electric field), and the class of molecule used (including how the molecule interacts with the contacts). This is illustrated by considering the energy level diagram in Figure 1C, which shows interfacial barriers for transport of holes ($\phi_{h+}$) or electrons ($\phi_{e-}$) at the interface between the conductive contacts and the molecular component of the device. The magnitude of the barrier is often correlated to the relative positions of the contact Fermi levels ($E_f$) and the frontier orbitals of the molecule (HOMO for holes, LUMO for electrons). Together with the total distance between the contacts, $\phi_{h+}$ and $\phi_{e-}$ are critical parameters in determining the mechanism and efficiency of charge transport in the device. However, in addition to these parameters, other details must be considered in order to appreciate all of the factors that control charge transport[10]. This requirement to consider the whole system has been discussed in previous studies that have demonstrated that perturbations of system energy levels can arises from a variety of interactions[11] such that the energy levels of the isolated components have limited predictive value[10]. This has led to a desire to develop techniques that characterize operating molecular junctions, so that information about the entire system can be characterized. One way in which to carry out such studies involves the interaction of light with molecular junctions.

A variety of optical characterizations on molecular junctions have been carried out. Several reports of Raman spectroscopy with simultaneous structural and electrical measurements[12-15] have shown that structural detail can affect the conductance of molecular junctions and that the bias voltage can induce structural changes in the molecular component. Ultraviolet-visible spectroscopy has been used to characterize the electronic energy levels of molecular layers used in electronic junctions, either to estimate the HOMO and LUMO levels of molecules[16, 17] or to follow dynamic redox reactions in devices under bias[18, 19]. Infrared spectroscopy has also been used to verify the structure of molecular layers after top contacts have been deposited[20, 21]. Other optical methods have also been employed in order to more directly probe transport (as opposed to chemical structural features). For example, photocurrent measurements have been used to determine interfacial barrier heights through the use of internal photoemission (IPE)[22-25].

The generation of photocurrent in molecular junctions has been considered theoretically by Galperin and Nitzan[26] and some experimental measurements of a range of photoeffects have been reported[27-29]. When studying photocurrents, it is important to distinguish between two different regimes- one in which the photocurrent originates in the contacts and one where carriers originate in the molecular layer itself. In the case of internal photoemission (IPE), as shown in Figure 1C, for incident photon energies where molecular absorption is weak (i.e. there is negligible light induced excitation of the molecular component)[30], charge carriers generated in the contacts can produce photocurrents that correlate to system energy levels in an intact, operating device. In the past, IPE has been used to characterize system energy levels (e.g., the value of the interfacial tunneling barrier). However, for incident photon energies for which the molecular component absorbs light, the basic character of the photocurrent contains different information. In this article, we consider a set of aromatic molecules that absorb light over some part of the photocurrent spectrum, and we analyze an extended range of layer thicknesses. By correlating the UV/vis absorbance of a range of molecular structures to the photocurrent spectrum, we gain insights into the nature of electronic coupling and electronic transport in nanoscale molecular devices.

## 2 EXPERIMENTAL

Figure 1A shows the schematic of the junction structure used in this study. The fabrication of molecular junctions has been described in numerous past reports[31-35], and began with the fabrication of flat carbon surfaces[36] by the pyrolysis of photoresist films. Briefly, photoresist was spin-coated onto polished quartz substrates and patterned using photolithography. Quartz substrates were rather than Si/SiOx both to avoid photoeffects from silicon and to permit UV-Vis absorption measurements. After developing 0.5 mm lines of photoresist, samples were pyrolyzed by heating to 1100 °C for 1 hour in forming gas (5% $H_2$ in $N_2$). Growth of molecular layers onto the carbon substrates utilized the electrochemical reduction of diazonium reagents using the carbon samples as the working electrode in a three-electrode cell (Pt wire auxiliary electrode and Ag/Ag+ reference electrode) in a dilute solution of the diazonium compound in acetonitrile (generally 1 mM concentration of diazonium reagent with 0.1 M tetrabutylammonium tetrafluoroborate as supporting electrolyte). Diazonium compounds were synthesized in-house through diazotization (isolated as tetrafluoroboro salts) of the corresponding aromatic amine compounds, where the location of the amine determined the location of the radical after electrochemical reduction (structures of the diazonium reagents used in this work are shown in Figure 1). Two molecules (Ferrocene and BTB) utilized an in-situ method for layer growth, as described elsewhere[1]. By controlling the parameters of the electrochemical voltage sweep program (i.e., the potential range, number of sweeps, and scan rate), the thickness of the films can be controlled in the range of 2-65 nm, with verification of thickness carried out using AFM scratching[35, 37]. For making completed molecular junctions, top contacts were deposited using electron-beam deposition of 20 nm of copper[31].

The diagram of the optical and electronic apparatus can be found elsewhere[24, 25]. Briefly, a small band ($\Delta\lambda$ = 13 nm) of light from a Xe arc source passed through an optical chopper before incidence onto the junction. Dual phase sensitive LIA detection was used to measure the resulting photocurrent, which was converted to yield by calibration of the incident power. Using this experimental setup, we measured the photocurrent response under ambient conditions at room temperature of molecular junctions composed of carbon/molecule/Cu with systematic variation of the molecular components, molecular layer thickness, and photon energy. The dark current was ~3 pA, which was used to determine the limit of detection (LOD) for photocurrent measurements. Each data point in the photocurrent yield spectrum represents the average value of 4 or 5 molecular junctions for each molecular structure and each thickness.

UV-Vis absorption spectroscopy was performed using a Perkin Elmer Lambda 1050 UV/Vis/NIR spectrometer with a photomultiplier detector. Scans were done in the wavelength range of 250-800 nm (corresponding to energy range 1.5 eV to 5 eV) with a resolution of 2 nm and an integration time of 1-2 seconds. Two types of absorption measurements were done: (a) dilute solutions of the monomer molecules and (b) molecular layers chemisorbed onto a transparent carbon surface. For solution measurements, monomers (as obtained from Sigma-Aldrich) of known concentration (0.1 to 0.7 mM) were dissolved in cyclohexane and the spectra were acquired, with a cyclohexane reference. For optical absorption measurements of chemisorbed films, the same procedure described above was used, but the photoresist was diluted to 5% (V/V) using propylene glycol methyl ether acetate, and no photopatterning was carried out prior to pyrolysis, as described previously[21, 25, 38]. Absorbance data were corrected by subtracting the background of the unmodified transparent substrate as described previously[39]. The LOD of the absorption measurements was determined as three times the standard deviation calculated from the flat part of the spectrum (where there is no observable absorption). Yield was calculated as described elsewhere[24].

## 3 RESULTS AND DISCUSSION

The characteristics of light absorption by molecular materials are known to undergo changes when the material is placed onto a conductive surface. For example, aromatic molecular layers chemisorbed to carbon electrodes using the reduction of diazonium compounds show a red-shifted absorbance maximum ($\lambda_{max}$), which can also be stated in terms of the energy of maximum (i.e. peak) absorbance, ($E_{max}$) relative to the monomer molecule in solution, along with a broadened peak shape[38]. Figure 2A shows the UV/vis absorption spectrum of a molecular layer of azobenzene (AB) in solution (bottom) and bonded to a PPF conductive substrate (top). Several important observations that can be made by analysis of these spectra. First, the peak locations are red-shifted after bonding to the surface (listed in Table 1). Second, the spectrum for the bonded AB shows a red-shifted tail, with significant absorption at lower energies compared to the solution spectrum. This shift is reflected in the absorption onset values, determined as shown in Figure 3 and listed in Table 1. In general, a red-shift is an indication of electronic delocalization, indicating that the molecule and the conductive substrate are coupled. Note that this also indicates that a molecular layer will show optical absorption at perhaps unexpected energies compared to the molecule in solution. Finally, there is significant broadening of the peaks for the chemisorbed molecular layer, which serves as an indication that the molecule-electrode system contains a distribution of states, some of which do not exist in the free molecule. When combined with the observed low-energy tail, the spectra show that states within the HOMO-LUMO gap of the free molecule exist in the molecular layer-electrode system, such that the optical measurements may probe a variety of states, some of which are relevant to electronic transport[40]. Figures 2B and 2C show similar plots for NAB and AQ, illustrating that the observations are consistent for a series of 10 molecules, with relevant observations listed in Table 1.

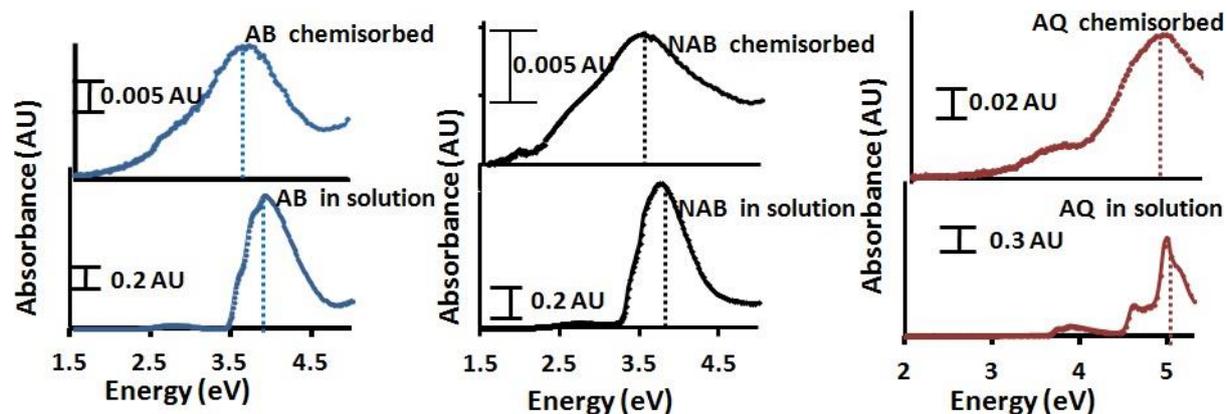

**Figure 2**. UV/vis absorption spectrum of molecules bonded to a transparent carbon substrate (top) and in cyclohexane solution (bottom) for (A) AB (B) NAB, and (C) AQ. Data for AB and NAB are new data, similar to that presented previously[25, 38].

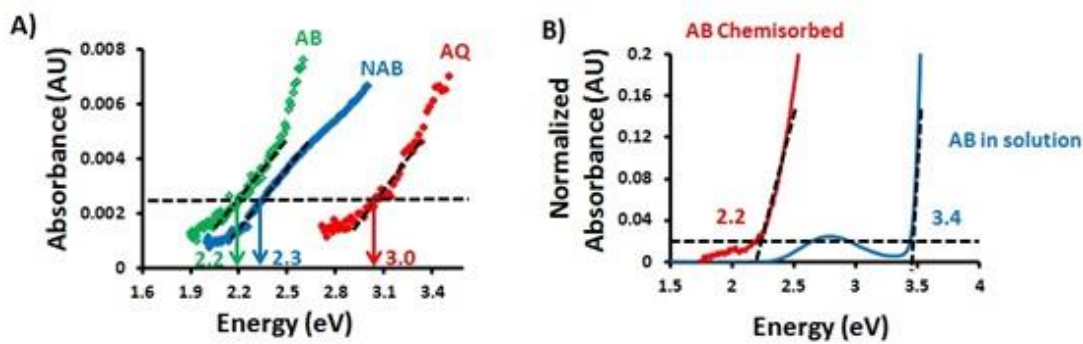

**Figure 3**. (A) Close-up of the UV/vis spectrum near the low-energy cut-off region for AB (green curve), NAB (blue curve), and AQ (red curve) chemisorbed on carbon plotted against the 3 sigma line, where the crossing point is defined as the low-energy absorbance on-set ($E_{onset}$, see Table 1). (B) Close-up of the absorption onsets for AB chemisorbed (red curve) and in solution (blue curve).

**Table 1**. Summary of UV/vis absorbance parameters for several aromatic molecules used in this study.

| Sample | Solution Onset (eV) | Chemisorbed Onset (eV) | Solution $E_{max}$ (eV) | Chemisorbed $E_{max}$ (eV) | Solution FWHM (eV) | Chemisorbed FWHM (eV) |
|---|---|---|---|---|---|---|
| Nitroazobenzene (NAB) | 2.3 | 2.3 | 3.8 | 3.5 | 0.75 | 2.5 |
| Azobenzene (AB) | 3.5 | 2.2 | 3.9 | 3.7 | 0.7 | 1.9 |
| Nitrobiphenyl (NBP) | 3.7 | 2.9 | 4.2 | 3.8 | 0.75 | 1.4 |
| Anthraquinone (AQ) | 3.5 | 3.1 | N/A | 4.8 | 0.35 | 1.3 |
| Nitrophenyl (NP) | 4.0 | 2.7 | 4.9 | 4.4 | 0.76 | 1.5 |
| Biphenyl (BP) | 4.4 | N/A | 5 | 4.6 | 0.7 | N/A |
| Bromophenyl (BrP) | 5.3 | 4.1 | >6 | >6 | N/A | N/A |
| Bisthienyl benzene (BTB) | N/A | 2.5 | N/A | N/A | N/A | 1.4 |
| NDI | N/A | 3.0 | N/A | 3.2/3.4 | N/A | 0.5 |
| Ferrocene (Fc) | N/A | 3.1 | N/A | 4.1 | N/A | 0.5 |

Analysis of Table 1 shows that a variety of molecules exhibit significant changes in absorption characteristics when the chemisorbed oligomer films are compared to the free monomers in solution. In particular, the bonded molecules have $E_{max}$ energies that are 0.2 to 0.5 eV lower in energy than the corresponding free molecule $E_{max}$, with the onset of absorbance as much as 1.3 eV lower in energy. The values for the FWHM of the main absorbance peaks shows a consistent broadening of 0.6 to over 1.5 eV compared to the monomers. These data serve as an indication that the states involved in optical absorption by the molecular component are fundamentally altered upon bonding to the substrate. As stated above, this can arise from new states in the molecular layer-electrode system.

In addition to interactions with substrates, it is known that oligomers can show shifts in $E_{max}$ that depend on the way in which monomer units are conjugated across the entire structure, with larger shifts to the red for larger electron delocalization[41]. These shifts in $E_{max}$ often vary inversely with length, with an approach to a limiting value defined by the monomer structure on one end, and the maximum extent of conjugation on the other (which is often termed the effective conjugation length, or ECL). Because diazonium growth is a radical-mediated growth process, a variety of possible structures can exist in an as-grown film[42]. While we will not be concerned here with the details of the radical growth mechanism, experimental measurements of the optical absorption properties of these films are reported in order to provide insights into the optical absorption characteristics in the as-grown films, with a consideration to how this parameter can influence photocurrent measurements discussed later.

Figure 4 illustrates the effect of molecular layer thickness on optical absorption by overlaying the measured absorbance for three different molecular structures as a function of layer thickness (A,C, E). The trend for the main peak absorbance value is also shown (B, D, F), illustrating that the absorbance follows a linear trend. Optical absorption ($A$) is often described using the Beer-Lambert law

$$A = \varepsilon bc \qquad (Equation\ 1)$$

where $\varepsilon$ is the extinction coefficient, $b$ is the path length, and $c$ is the concentration of the absorbing species. For a solid organic film, the thickness of the molecular layer can be taken as $b$, and the packing density (i.e., the number of molecules per $cm^2$) can be taken as $c$. Thus, a linear increase in $A$ with $d$ (as reported in Figure 4A and B) implies that the packing density is constant as the layer thickness increases.

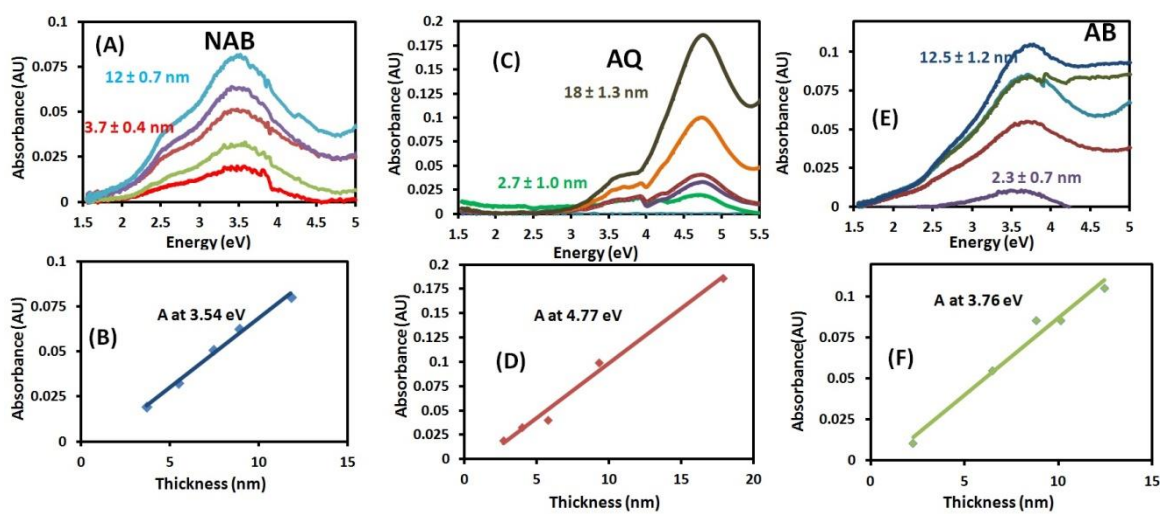

**Figure 4**. Optical absorption spectra and $E_{max}$ as a function of thickness for NAB (A and B), AQ (B and D) and AB (E and F). The values of Emax are given in Table 1 and in each Figure.

An additional notable feature in Figure 4 is that $E_{max}$ does not shift significantly in energy as thickness increases, indicating that there is little change in the extent of electronic coupling within the molecular layer. Computations are underway to determine the possible causes for this effect, but we note here that the constancy of $E_{max}$ means that the amount of light absorbed at $E_{max}$ is expected to follow Equation (1). Because it is also demonstrated in Figure 4 that the spectral shapes (i.e., the peak shapes and positions, along with the low-energy onset of absorption) do not change significantly with thickness in the ranges tested, measurement of the photocurrent is expected to reflect the overall intensity of absorption at a particular energy for each structure, as elaborated below. Figure 4 shows that the experimental $E_{max}$ values do not change with thickness, but that they do vary with molecular structure.

When light is incident onto a molecular junction, there are different mechanisms for photocurrent generation that can be classified based on where the photon-charge carrier conversion takes place. Besides bolometric photocurrent (which was discussed and eliminated previously[24, 25]), there are two broad mechanisms for generating photocurrent to consider: internal photoemission (IPE) and molecular absorption. IPE can be understood by considering that external photoemission, also known as the photoelectric effect, occurs when carriers are given enough energy by incident photons to escape the conductor (i.e., when the photon energy exceeds the material work function). However, when there are internal system barriers in a conductor-insulator-conductor system, a photocurrent can result when the incident photon energy exceeds an interfacial barrier, thus allowing carriers to move. Such photocurrents can occur when the internal system barrier is much lower than the contact work function(s), as is most often the case. Thus, IPE originates through absorption of photons in a contact, which can excite plasmons that subsequently decay to generate hot carriers[43], resulting in photocurrent that can be used to measure interfacial barriers heights[44]. A schematic representation of the IPE process is shown in Figure 1C, represented using an energy level diagram, with molecular HOMO and LUMO levels offset from the contact Fermi levels ($E_f$). The numerical value of these offsets is often used to define the non-resonant tunneling barrier for the transport of holes ($\phi_{h+}$) or electrons ($\phi_{e-}$). As shown, the IPE process begins with photon absorption in the Cu contact, which leads to the generation of hot carriers that have a maximal excess energy defined by the photon wavelength (for electrons, this energy is $E_f + h\nu$, while for holes, it is $E_f - h\nu$). These photoexcited carriers can cross internal system barriers when $h\nu$ exceeds $\phi_{h+}$ or $\phi_{e-}$. However, the magnitude of the photocurrent will depend on several factors, such as the intensity of the incident light, as well as the total thickness of the molecular layer separating the contacts. Here, we vary the thickness of the layer in order to determine the effective distance over which IPE can be maintained.

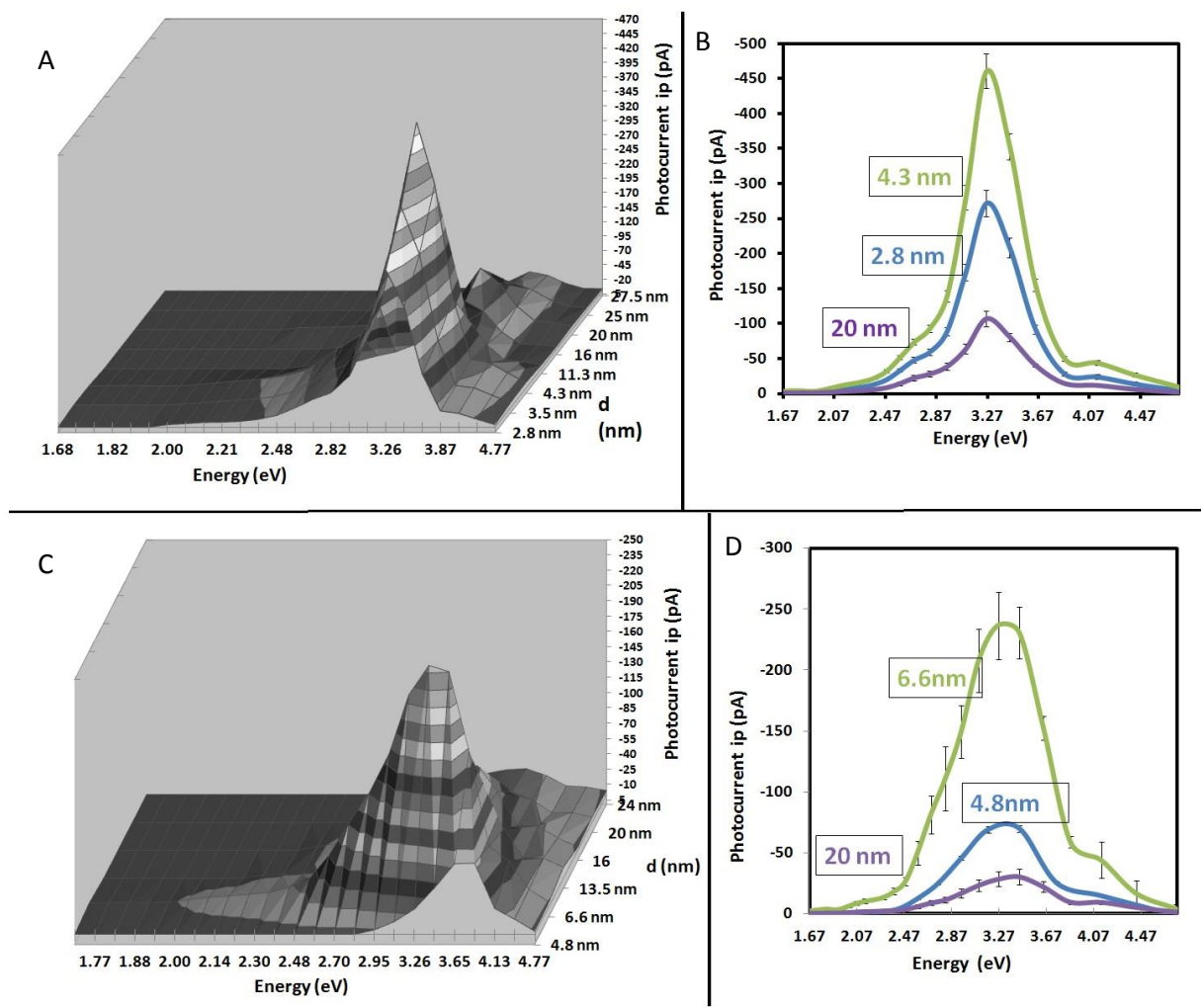

**Figure 5.** 3-D plot of photocurrent as a function of energy and thickness for (A) NDI along with a cross-section of data for three thicknesses. Similar 3-D plot for AQ (C) with a cross section (D). In (A) and (C), the low-lying black areas represent no detectable photocurrent, illustrating that for thin molecular layers, photocurrent is observed that tends to diminish as thickness increases.

Figure 5 shows photocurrent spectra for two series of molecular junctions, NDI and AQ, with the added dimension of thickness. In Figure 5A, the black shading indicates no measureable photocurrent. Thus, for thick layers, at low photon energies (the top left of the plots), the photocurrent decays as the absorption onset is approached. Near the absorption peak, on the other hand, the photocurrent shows a strong signal. The boundaries of these different regions are apparent in Figure 5, with large peaks in in photocurrent for both cases at the absorbance maximum and 5-7 nm in thickness. The location of these peaks and their fall-off with increasing thickness suggest that an important limit is reached near 10 nm, after which the photocurrent decreases with greater thickness. In addition, these plots show a clear continuation of measureable photocurrent to over 20 nm, although the relationship between the photocurrent and thickness becomes a complex function (as discussed in more detail below). Figure 5B shows the representative photocurrent spectra for 3 different thicknesses of 2.8nm (smallest thickness), 4.3nm and 20nm (highest thickness) from the series of NDI MJs. We clearly see the photocurrent intensity increases from 2.8nm to 4.3nm and there is a marked decrease in photocurrent intensity for 20nm NDI molecular junctions. This was also observed in the series of AQ molecular junctions as illustrated in Figures 5C and D.

As observed in Figures 5, where the chemisorbed molecular layer shows negligible absorption (at low energy), the value of $Y$ decreases below the detection threshold by 7-12 nm in thickness (i.e., the flat region in the upper left of the 3-D plot). The observation that the IPE-based photocurrent is extinguished by ~12 nm indicates that photoexcited carriers generated in a contact have limited range. This distance scale may be determined by a variety of phenomena, including scattering of carriers in the contact and molecular layer, the energy levels of the molecular orbitals that mediate transport relative to the contact Fermi level(s), and the magnitude of any built-in fields. In any case, this measurement provides an important indicator of electronic transport across molecular layers, and shows directly that there is a limited distance over which carriers can be transported without applied bias. This distance may be correlated with coherence, electronic coupling, or other fundamental distances in molecular devices.

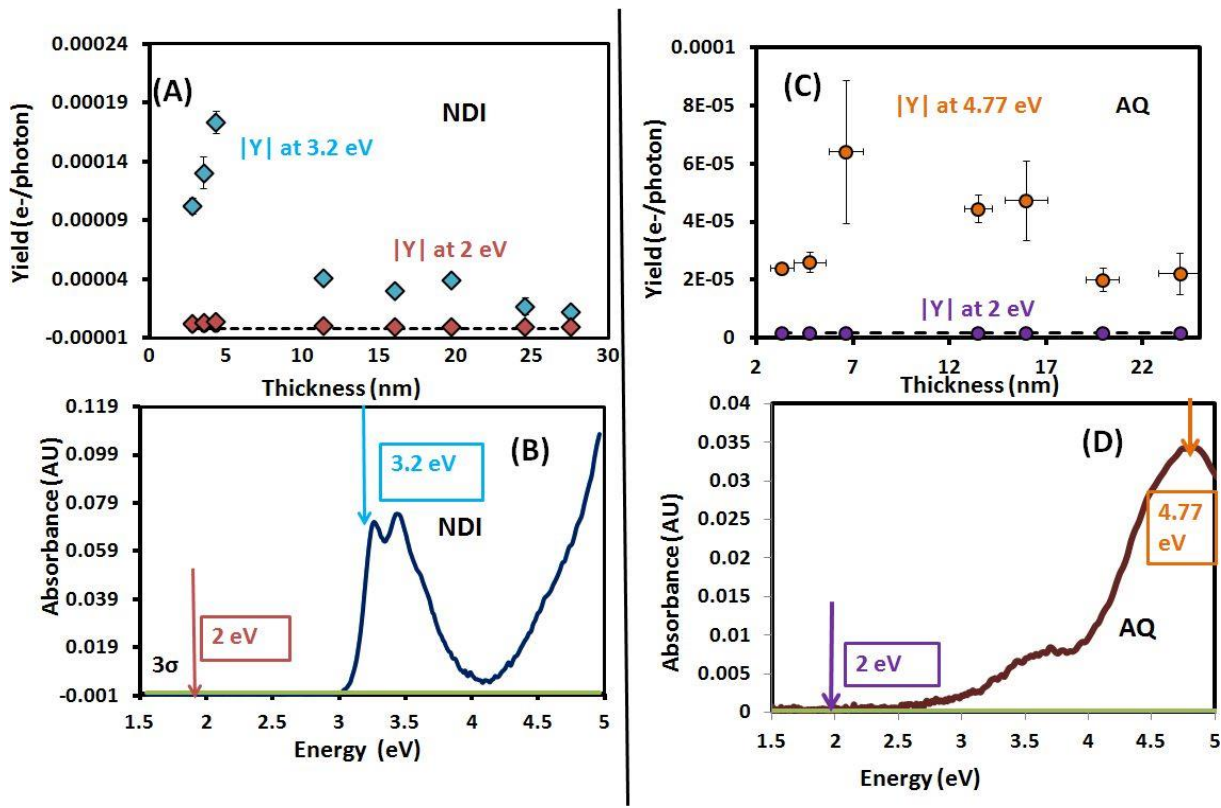

Figure 6 (A) Photocurrent yield vs. molecular layer thickness for NDI at two energies: 3.2 eV (blue diamonds) and 2.0 eV (red diamonds). The horizontal dashed line represents the limit of detection for the photocurrent measurement. (B) Absorption spectrum of NDI, showing that 3.2 eV is a characteristic absorption peak, while absorption is below the 3σ line at 2.0 eV. (C) Photocurrent yield for AQ devices as a function of thickness for two energies, 2.0 eV and 4.4 eV. (D) Absorption spectrum for a molecular layer of AQ, showing no absorption (below LOD) at 2.0 eV, and strong absorption at 4.4 eV.

Figure 6 shows a comparison of the photocurrent yield at two different photon energies versus thickness for NDI and AQ along with the absorption characteristics of spectra of the chemisorbed molecules. The two relevant energy values considered are at or near $E_{max}$ and 2 eV, where molecular absorption is undetectable in each case. Importantly, the yield recorded at or near $E_{max}$ shows a distinctly different behaviour compared to the yield at an energy where no the absorption takes place. This contrasting behaviour demonstrates that the physical processes taking place in the molecular junction that produce the measured photocurrent are quite different for the two different energy regimes. For example, at low energy, there is no measurable photocurrent for NDI or AQ devices for $d > 7$ nm, while at or near $E_{max}$, photocurrent is easily measured for $d = 20$ nm and beyond.

Additional analysis of other cases at $E_{max}$, as presented below, show some trends that provide insights into this behaviour.

Figure 7A shows photocurrent yield at the energy of maximum optical absorbance for several different molecular junctions containing a range of thicknesses from 2-27 nm. For both NDI and AQ, the photocurrents are much larger when the incident light is within the absorption band of the molecular layer. Interestingly, in each case there is a linear or near-linear increase of yield with thicknesses less than 7 nm, after which the yield values begin to decrease. This region is shown in detail in Figure 7B, with linear regression fits as shown. These data also show that photocurrent can be measured for $d$ up to at least 27 nm, indicating that IPE is not the only mechanism in operation, since carriers generated in the contact do not result in photocurrent beyond ~7 nm, as discussed above. Taken together with the results from Figure 4, the results in Figure 7 indicate that while the molecular layer increases the amount of light that is absorbed as the layer thickness increases, the photocurrent does not follow the same trend, with linearity being lost above ~6 nm.

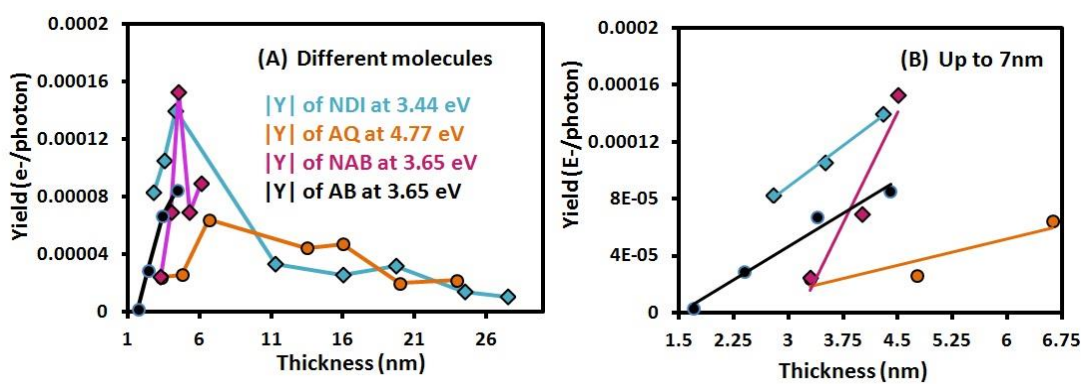

**Figure 7.** (A) Photocurrent yield versus thickness for four different molecular structures as a function of thickness at the maximum absorption energy. (B) Close up of the region below 7 nm, where data shows a linear trend. The datapoints in (B) follows the same color scheme as in (A)

Considering all the results, schematic energy levels and transitions underlying IPE and MA are proposed in figure 8. For IPE (8A) there is a photocurrent due to excitation of carriers in the top contact which traverse the molecular layer provided the layer thickness is less than ~10 nm. For thicker layers, scattering and other mechanisms lead to loss of energy and no photocurrent for the IPE mechanism beyond ~10 nm. It is important to note that these experiments were carried out at zero applied bias. It may be possible to extend the range for IPE by applying an electric field, and this could extend the insights possible from photocurrent experiments. However, to remain strictly in the IPE regime, light absorption by the molecule should be negligible, as it obviously complicates the photocurrent response.

Figure 8B shows a proposed mechanism for photocurrent generation when the molecular layer absorbs light. As shown, there are excitons created throughout the film, both near the interfaces and in the interior of the molecular layer. In addition to recombination, there are at least two possible events which lead to charge separation and photocurrent, and these depend in part on the location of the exciton relative to the electrode surfaces. First, an electron or hole may be transferred directly to the electrode if the exciton is close enough for direct electronic communication. Second, the exciton may diffuse toward an electrode, then undergo charge separation. For molecular junctions where the total thickness of the molecular layer exceeds the range of direct of electron transfer, direct transfer is possible only for a fraction of the excitons generated in the layer, leading to a deviation of the photocurrent yield versus thickness from a monotonic increase for thin molecular layers to a more complex behaviour. While there are many pathways for the photocurrent trend to deviate from increasing monotonically with thickness, most involve processes that depend on distance. For example, a direct transfer of charge from the molecular layer to either contact cannot take place if the exciton is not electronically coupled to that contact. The distance over which electronic coupling can extend is limited, and when the total thickness of the molecular layer exceeds this length, a variety of mechanisms can cause dissipation of the energy, including recombination of bimolecular or unimolecular character. In our previous work[25], we showed that when the molecule absorbs light, we observed that photocurrent increased with the square root of the incident light intensity, a trend consistent with bimolecular recombination. While the intent here is not to provide an analysis of specific recombination mechanisms, we reiterate that the importance of the results is in showing that below 6-10 nm, there is an indication of direct electronic communication between the molecular layer and the contact(s). Thus, the recombination or dissipation shown in Figure 8 could result in reduced photocurrent and deviation from the linear increase in photocurrent with layer thickness apparent in Figure 8B.

We also note some important features of the mechanism proposed in Figure 8B that are not readily apparent from a schematic depiction. First, we again emphasize that these experiments are carried out at zero bias, such that no net driving force is externally imposed on the system. Carrying out the experiment under bias may yield additional insights, such as how the electric field influences the diffusion of carriers at the nanoscale. Second, the direct transfer of charge carriers at the interface that is implied by the linearity of the photocurrent for thin molecular layers implies that transport across the interface is rapid. As long as the rate of transport across the interface is more rapid than transport of excitons through the film, the measurement of photocurrent in molecular devices for energies where the molecular layer absorb strongly will provide an indication of the electronic coupling length between the molecular component and the contact. The point at which the photocurrent diverges from the linear trend observed in absorbance may be the escape radius[45] of the charge carriers, a concept that could be correlated to the electronic coupling length and strength between the molecular layer and the contact(s).

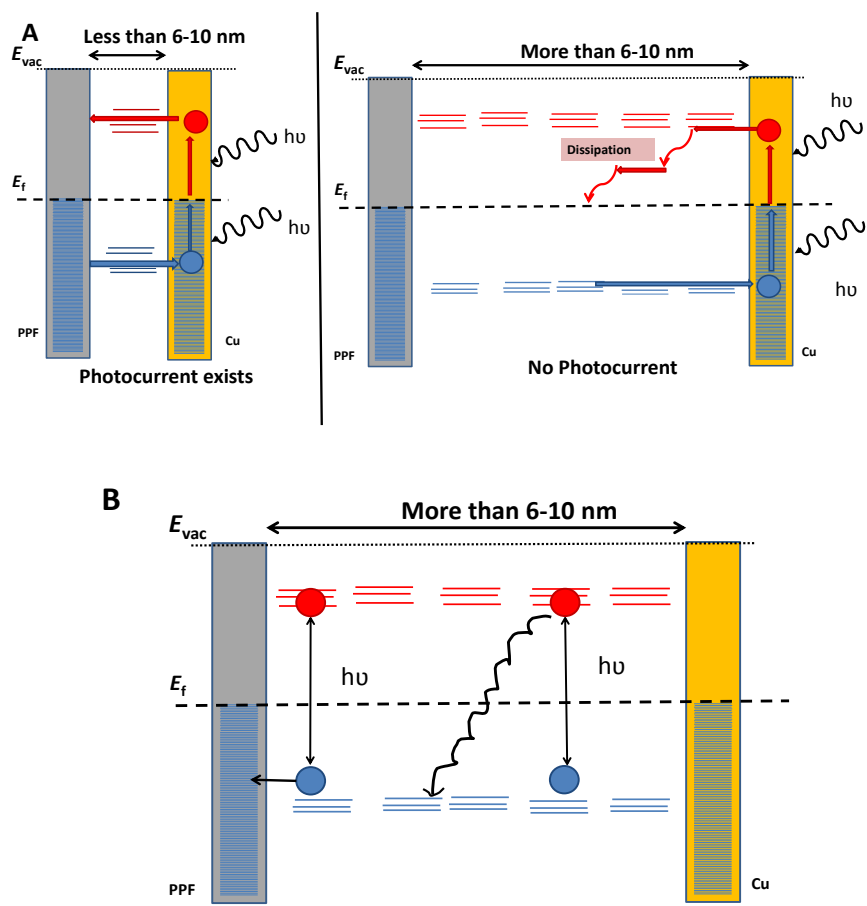

**Figure 8**. (A) Left side shows IPE producing measurable photocurrent up to ~10 nm. Right shows thicker layers that do not show measurable photocurrent due to various hot carrier dissipation mechanisms. (B) Schematic of the molecular absorption photocurrent mechanism, illustrating that a direct charge separation can only take place within a limited distance, observed from our experiments to be ~6-10 nm.

## 4 CONCLUSIONS

This paper has demonstrated that for the observed photocurrent spectrum of aromatic tunnel junctions, there are two distinct regimes, one involving IPE and a second where light induced excitation of the molecular component takes place. It has been demonstrated that for energies where the molecular layer does not absorb light (but where the Cu top contact does), IPE can generate measurable photocurrents for junctions no thicker than ~10 nm. in distance, presumably due to scattering of carriers or other types of losses associated with hot carriers traversing a barrier region. In order to explain photocurrents beyond distances where IPE can operate, it is clear that absorption by the molecular layer is important, leading to an investigation of absorption as a function of thickness for three different structures. Photocurrents resulting from optical absorption by the molecular layer are much larger than those from IPE, and were observed for layer thicknesses up to 27 nm. It was shown that coupling of electronic states between the molecule and electrode (i.e., molecule-electrode hybrid orbitals) is present, but that this coupling (as measured by the maximum absorption energy and broadening of the optical absorption band) does not change with increasing film thickness. Thus, while optical absorption increases approximately linearly with molecular layer thickness, the photocurrent yield deviates from this trend, possibly due to the different length scales involved in absorption and charge transport. By advancing our understanding of the absorption-mediated photocurrent in these devices, important length scales involved in charge transport may be investigated directly in an operating device, and in the present case, a distance of 5-7 nm appears to be the maximum electronic coupling distance, where electronic communication between excitons in the molecular layer and the contacts is established. This result will be important in considering electronic transport mechanisms beyond this important distance scale in molecular electronics.

## 5 ACKNOWLEDGMENTS

This work was supported by the University of Alberta, the National Research Council of Canada, NanoBridge, and Alberta Innovates. We thank Bryan Szeto for expert assistance with LabView programming, and Mykola Kondratenko for fabrication of NDI and Fc devices.